# DynaPain: Moving Flame Beetle with Dynamic Pain Illusion Adapting Apparent Movement to Thermal Grill Illusion

Souta Mizuno[1], Jiayi Xu[2], Shoichi Hasegawa[3], Naoto Ienaga[4], and Yoshihiro Kuroda[4]

[1] Graduate School of Science and Technology, University of Tsukuba, Ibaraki, Japan

[2] Research Center for Advanced Science and Technology, The University of Tokyo, Tokyo, Japan

[3] Laboratory for Future Interdisciplinary Research of Science and Technology, Tokyo Institute of Technology, Tokyo, Japan

[4] Institute of Systems and Information Engineering, University of Tsukuba, Ibaraki, Japan

(Email: mizuno@lelab.jp )

**Abstract ---Pain sensation presentation with movable sensory position is important to imitate the pain caused by objects in motion and the pain corresponding to a person's movements. We aimed at proposing a novel dynamic pain sensation experience, called DynaPain. DynaPain was achieved by the non-contact thermal grill illusion and the apparent movement. The demonstration provided the dynamic heat and pain experience through interaction with a flame beetle moving on the arm.**

## 1 Introduction

Pain is an important element in human perception that detects danger. Sensory presentation that deals with pain has attracted attention in Virtual Reality (VR) and Augmented Reality. The sensory presentation of pain is expected to be applied to training for dangerous situations and exposure therapy. Conventionally, electrical stimulation [1] and Thermal Grill Illusion (TGI) [2], which is an illusion using thermal stimulus, have been used for the sensory presentation of pain. TGI is a high-intensity and non-invasive stimulus that is useful for presenting imitative sensation in VR. However, existing TGI presentation methods are mainly static sensory presentations. Therefore, sensory presentation with motion is difficult due to factors such as the physical movement of the device and the time required to perceive the illusion. In this study, we propose a novel pain sensation experience that includes a moving pain illusion. Examples of when a dynamic pain illusion is required include training on machine tools and entertainment experiences that involve moving stimuli. The representation of moving pain stimuli provides a rich experience that involves impactful or invasive interaction in a virtual environment. We developed DynaPain, a movable pain illusion presentation system, by generating apparent movement through TGI switching. DynaPain is constructed by the non-contact TGI presentation system using cold air flow and optical heating.

## 2 Method

### 2.1 Non-Contact TGI Presentation

The non-contact TGI presentation method uses cold air flow and optical heating with Light Emitting Diode (LED) light. Cold air flow is provided by a vortex tube that utilizes the vortex effect [3] to divide air into cold and warm air. LED light is superior in terms of immediacy of response, as it can quickly present heat and switch on or off. The intensity can be adjusted from the flow rate and duty ratio by PWM control. In addition, there is no heat transfer due to remaining heat of the element itself. By using these elements to form a warm and cool distribution, the TGI presentation can be immediately switched between presenting and not presenting.

## 2.2 Apparent Movement for TGI

As a dynamic presentation method for TGI using the illusion of movement. We focused on apparent movement [4], in which a sense of movement is perceived from two different stimuli. This is an illusion of movement between stimuli, which enables the generation of a sense of movement that occurs when the stimulus is moved. To present this illusion, a non-contact TGI device is used to switch between two stimuli without residual heat, and to generate apparent movement of the TGI. This generates a dynamic pain illusion.

## 2.3 Device and Presentation

The presentation device uses a cold air flow port, LEDs, and a slit to form a warm and cool distribution. In this experiment, the Arduino is operated at 115200 bps to send signals to LEDs and solenoid valves for cold air flow control. A temperature change of about maximum 42 degrees Celsius for the LED and about 26 degrees Celsius for the cold-air flow is presented at the time of presentation. In order to avoid the perception of heat-induced pain in humans, the upper limit for the LED thermal stimulus was set at about 40 degrees Celsius in this study. The apparent movement response of TGI with increasing LED intensity showed a tendency to change the optimal location at each intensity, but this is still unclear.

## 3 Demonstration

Fig. 1 shows the DynaPain system. Non-Contact TGI presentation and the generation of apparent movement are controlled by switching the cold air flow and LED presentation. This realizes the presentation of a dynamic pain illusion. Fig. 2 shows a demonstration overview. Demonstration is a dynamic pain experience system. The stimuli in response to the insect-like character's "Flame Beetle" movement are presented to the users. Then, moving stimuli, changes in sensation of movement, TGI, and thermal feedback are presented to the users. Through these presentations, DynaPain provides a VR experience

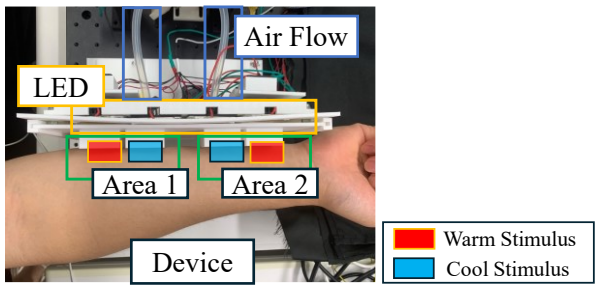

**Fig. 2  System Configuration**

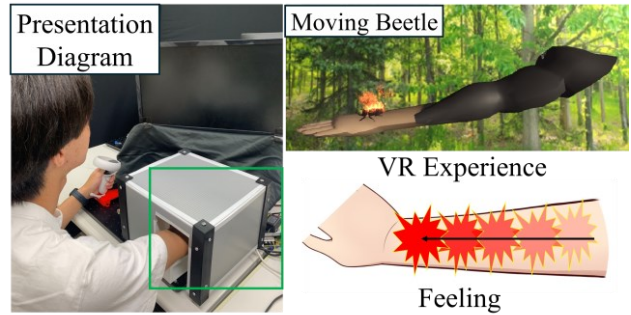

**Fig. 1  Demonstration** of dynamic pain sensation.

## 4 Conclusion

We have developed a non-contact pain illusion device as a pain sensation presentation method for imitative sensation presentation. Then, we realized a dynamic pain sensation presentation by combining apparent movement as a more realistic sensation presentation. As an experience using this, we created a demonstration of a moving stimulus source with a matching dynamic pain presentation. We hope that this research will contribute to the development of a new sensory presentation method in VR and will be useful for future sensory presentation.

### Acknowledgement

This research was supported by JSPS Grants-in-Aid for Scientific Research JP21H03474, JP24K02969, and JP24K22316.